\documentclass[aps]{revtex4}
\usepackage{amsfonts}
\usepackage{graphicx}
\usepackage{pifont}
\usepackage{graphicx}
\usepackage{amsmath,amssymb,amsthm}

\newcommand{\ini}{\begin{equation}}
\newcommand{\fin}{\end{equation}}
\newcommand{\inia}{\begin{eqnarray}}
\newcommand{\fina}{\end{eqnarray}}

\begin{document}

\title{\textbf{Localized Gravity on Topological Abelian Higgs Strings.}}
\author{Rafael S. Torrealba\ S. }
\email{rtorre@ucla.edu.ve}
\affiliation{Departamento de F\'{\i}sica. Universidad Centro Occidental Lisandro Alvarado.}

\begin{abstract}
In this work there are proposed several new 6D brane worlds based on exact
topological solutions to the vortex equations for the Abelian Higgs model,
with suitable boundary conditions and symmetry breaking potentials similar
to the mexican hat. Just as happens for RS domain wall brane worlds, it is
shown that the massless mode of linearized gravity is localized in the
neighborhood of the string-vortex solution while there are not massive
bounded states. The correction to the newtonian limit of the effective 4D
gravity is calculated.

\vspace{0.5 cm} PACS numbers: 04.20.-q, 11.27.+d, 04.50.+h
\end{abstract}

\maketitle



\vspace{0.3cm}


\section{Introduction}


Brane worlds theories on domain walls had achieved successful development in
recent years, specially after the important results by Randall and Sundrum
explaining the hierarchy problem in a model with two branes in an AdS
orbifold \cite{RS1}, and the confinement of spin 2 massless gravity with the
classical newtonian limit \cite{RS2} in one brane model. Most of the
development on brane world theory as confinement of chiral fermions,
localization of gauge fields, BPS solutions, wide and thin domain walls
limits etc. had been achieved on 5 dimensional domain walls \cite%
{ReferencesDW}. Domain walls are not free of problems, the lost of energy by
radiation, the stability under casimir energy and/or quantum fluctuation,
the localization of gauge field, the supersymmetric extension and its
relation with inflation and are still unsolved problems currently under
research \cite{StabilityDW}.

Brane worlds also had been proposed on 6 dimensional string-vortices \cite%
{RSVortex} and higher dimensional topological defects \cite{RSmonopole}\cite%
{Bezerra}\cite{Hedgehog}. It is natural to expect better phenomenological
behavior with more extra non compact dimensions, in the same way that
smaller observable size is obtained for larger number of dimensions in AADD
models \cite{ADD} or greater warping volume factor arises with a greater
number of warped dimensions in RS models \cite{RS1}\cite{RS2}. Besides
domain walls, vortices are the simplest topological defect. The
string-vortex defects are topological charged solutions to the equations of
the Abelian-Higgs model, with non trivial first homotopy group for the
vacuum manifold, \ i.e. there exist strings solutions (or holes in 2
dimensions) around which any loop will have non trivial winding number. Some
important advances as localization of massless gravitons \cite{Hedgehog}\cite%
{Shaposhnikov}, confinement of fermions \cite{RD}, decoupling of
graviphotons and graviscalar \cite{Giovannini1} and confinement of
electromagnetic spin 1 gauge fields \cite{Giovannini2}\cite{FluctuationBW}
had been recently achieved for gravitational string-vortex in 6D. Almost all
the works relays on numerical \cite{RSVortex}\cite{Giovannini1}\cite%
{Giovannini2} or asymptotical approximations \cite{Hedgehog}\cite%
{Shaposhnikov}\cite{FluctuationBW}\cite{GaugeinVortex}, because there are
not known general analytic solutions for curved 6D string-vortices in curved
space time, nevertheless RS scenarios based on asymptotical (and numerical)
solutions had been reported \cite{RSmonopole}\cite{Bezerra}\cite{RSVortex}.
In order to proceed to future developments as quantization and stability
under fluctuations studies it is important to obtain exact solutions. So the
principal purpose of this paper, is to show new exact solutions to
string-vortices on which a RS scenario could be constructed and gravitation
localized.

In section II we obtain the exact solution of the Einstein equations in 6
dimensions for 3 new symmetry breaking potentials: two of which are non
polynomial, and one resembles the classical quartic mexican hat. In section
III we give two different proofs of the localization of the zero massless
mode for linearized gravity fluctuations and also prove that massive modes
are not bounded to the vortex. In section IV we obtain the massive exact
solutions for linearized gravity fluctuations, and calculate the newtonian
limit for gravity. In section V we make same remarks and conclusions.


\section{New topological einstein abelian string-vortex solutions.}


We start from the 6 dimensional action%
\begin{equation}
S=-\int dx^{6}\sqrt{-G}\left[ \frac{{\ 1}}{{\ 2\chi }}R+\Lambda \right]
+\int dx^{6}\sqrt{-G}\left[ \frac{{\ 1}}{{\ 2}}(D^{A}\phi )^{\ast
}(D_{A}\phi )-\frac{{\ 1}}{{\ 4}}F_{AB}F^{AB}-V\right] ,  \label{EinsteinAH}
\end{equation}%
where the first integral is the 6D Einstein Hilbert action, with a bulk
cosmological constant $\Lambda .$In what follows we will set $\ \chi =\frac{%
8\pi }{(M_{6})^{4}}=1,$ where $M_{6}$ \ is the 6D Planck mass and we will
follow the notation in \cite{RSVortex}. Latin uppercase indices run over 6
dimensions, Greek indices run over 3+1 dimensional spacetime.

Here we will look for a geometry 2+3+1 composed by a 3-brane that contains
the 3+1 physical universe and 2 extra dimensions on which we can choose
coordinates $(r,\theta )$ and the metric is given by:%
\begin{equation}
ds^{2}=M^{2}(r)g_{\mu \nu }dx^{\mu }dx^{\nu }-L^{2}(r)d\theta ^{2}-dr^{2},
\label{metric}
\end{equation}%
with $g_{\mu \nu }=\eta _{\mu \nu }=(+,-,-,-)$. In this context $r$ and $%
\theta $ are the coordinates of the extra dimension, $L(r)$ acts as a radial
factor and $M(r)$ is a warp factor, rapidly decreasing when moving away from
a 4 dimensional 3-brane located at $r=0$.

Assuming that the scalar and gauge fields in (\ref{EinsteinAH}) depends on
the extra coordinates as in the Nielsen Olesen ansatz \cite{Nielsen-Olesen}%
\cite{Shaposhnikov}

\begin{eqnarray}
\phi (r,\theta ) &=&f(r).e^{in\theta },\text{ \ \ \ \ \ \ \ \ \ \ }\ \ n\in
\mathbb{Z}  \label{F} \\
A_{\theta }(r,\theta ) &=&\ \frac{n.\alpha (r)}{e},\text{\ \ \ \ }
\label{Atheta} \\
A_{r}(r,\theta ) &=&0,  \label{Ar} \\
A_{\mu } &=&0\text{ ,\ \ }  \label{Amu}
\end{eqnarray}%
we get the curved version of Nielsen Olesen vortex equations \cite{RSVortex}%
:
\begin{eqnarray}
\frac{d^{2}f}{dr^{2}}+(4m+l)\frac{df}{dr}-\frac{P^{2}f}{L^{2}}-\frac{dV}{df}
&=&0,  \label{CurvedVotexA} \\
\frac{d^{2}P}{dr^{2}}-(4m+l)\frac{dP}{dr}-e^{2}f^{2}P &=&0,
\label{CurvedVortexB}
\end{eqnarray}%
where
\begin{equation*}
P=n\ [1-\alpha (r)],\text{\ \ \ \ \ \ \ \ \ }m=\frac{d}{dr}\ln [M(r)],\text{
\ \ \ \ \ \ \ \ }l=\frac{d}{dr}\ln [L(r)].
\end{equation*}

String-vortex solutions are classically obtained with the boundary
conditions:

\begin{equation}
\begin{array}{cc}
f(r\rightarrow 0)=0, & f(r\rightarrow \infty )=1, \\
\alpha (r\rightarrow 0)=0, & \alpha (r\rightarrow \infty )=1,%
\end{array}
\label{Boundary}
\end{equation}%
by numerics and asymptotical methods. Very few solutions to the equations (%
\ref{CurvedVotexA}) and (\ref{CurvedVortexB}) are known. Indeed only the
Bogomoln'yi solution for the critical case \cite{Bogomoln'yi} \cite{Vega} is
known to be exact. Recently in \cite{belga} the boundary condition ($\alpha
(r)=1$) was used to find new exact solutions, so we will assume here:%
\begin{equation}
P(r)\equiv 0\text{ \ }\ \ \forall \text{\ \ \ }\ r\in (0,\infty )  \label{P}
\end{equation}%
and the equation (\ref{CurvedVortexB}) will be automatically satisfied.

Boundary condition (\ref{P}) lead to a apparent singularity at $r=0$ in the
gauge field when written in vectorial form \cite{Preskill}:

\begin{equation}
\overrightarrow{A}=-\frac{n}{er}\alpha (r)\widehat{u}_{\theta },
\label{vector}
\end{equation}%
that correspond to the 1-form%
\begin{eqnarray*}
A &=&A_{B}dx^{B}=A_{\theta }d\theta =-\overrightarrow{A}.\overrightarrow{ds},
\\
\text{with \ \ \ \ \ \ \ \ \ \ \ \ \ \ \ \ }\overrightarrow{ds} &=&rd\theta
\widehat{u}_{\theta },
\end{eqnarray*}%
where the minus sign comes from the signature of flat space coordinates, and
the arrow labels only 2D vectors because we have set (\ref{Amu}).

The model (\ref{EinsteinAH}) is invariant under the group U(1) of local
gauge transformation:%
\begin{equation*}
\phi \rightarrow e^{i\vartheta (x)}\phi \qquad \qquad A_{B}=A_{B}+\frac{1}{e}%
\partial _{B}\vartheta (x)
\end{equation*}%
So, when the phase of the scalar field is choose as
\begin{equation*}
\vartheta (x,r,\theta )=n.\theta \qquad \Longrightarrow \qquad \delta
A_{\theta }=\ \frac{n}{e}
\end{equation*}
\ that is exactly (\ref{Atheta}) when $\alpha (r)=1$, so the equation (\ref%
{vector}) is pure gauge and the singularity is in a non physical sector.

This could also been seen from the classical electromagnetic\ field:
\begin{equation*}
F_{AB}=0\qquad \text{or equivalently }\qquad \overrightarrow{E}=0\qquad
\text{and\qquad }\overrightarrow{B}=0
\end{equation*}%
Although this type of string-vortex has nor electric neither magnetic
charge, it is a topological solution, because it still has a non trivial
integer winding number

\begin{equation}
n=\frac{e}{2\pi }\int_{C}A,\qquad n\in
\mathbb{Z}
\label{homotopy1}
\end{equation}%
where $C$ is any closed curve around a "string" at $r=0$.

Although the electric and magnetic fields are zero, it is not possible to
continuosly pass or deforms a curve with a particular value $n$ into a curve
with a different $n$. For each homology class of curves $C$ we will get the
same integer value for $n$. Each different value $n$ will correspond to a
class of curves $C$ and labels an specific\ class of topological vacuum.

This kind of string-vortex solutions will be referred as "topological
string-vortices" and they are interesting because 6 dimensional brane world
could be constructed on it. In order to obtain a feasible Randall Sundrum
scenario in 6 dimensions, we still have to prove that that the metric factor
$M^{2}(r) $ is warped.

Einstein equations are obtained performing variations of the metric in the
action (\ref{EinsteinAH}) as in \cite{RSVortex} and \cite{Shaposhnikov}.

\begin{eqnarray}
l^{\prime }+3m^{\prime }+l^{2}+6m^{2}+3ml &=&-\tau _{o},  \label{taucero} \\
4m^{\prime }+10m^{2} &=&-\tau _{\theta },  \label{tautheta} \\
2ml+3m^{2} &=&-\frac{{\ 1}}{{\ 2}}\tau _{r},  \label{taurho}
\end{eqnarray}%
where $^{\prime }$stands for $\frac{d}{dr}$, and $\tau _{o},$ $\tau _{\theta
},$ $\tau _{r}$ are the non vanishing components of the energy momentum
tensor:%
\begin{eqnarray}
\tau _{o} &=&T_{0}^{0}=\frac{(f^{\prime })^{2}}{2}+V(f),  \notag \\
\tau _{\theta } &=&T_{\theta }^{\theta }=\frac{(f^{\prime })^{2}}{2}+V(f),
\notag \\
\tau _{r} &=&T_{r}^{r}=-\frac{(f^{\prime })^{2}}{2}+V(f),  \label{Trho}
\end{eqnarray}%
here (\ref{P}) was inserted into the equations and the 6D (bulk)
cosmological constant was absorbed into the redefinition of the potential $V=%
\tilde{V}+\Lambda $.

The equation for the scalar field (\ref{CurvedVotexA}) could also be
obtained, as in \cite{Bezerra}\cite{Shaposhnikov}, as a consequence of the
conservation of energy momentum tensor, by means of:%
\begin{equation}
\frac{d}{dr}\tau _{r}=4m(\tau _{o}-\tau _{r})+l(\tau _{\theta }-\tau _{r}),
\label{conservation}
\end{equation}%
so, given a potential $V(f)$, equation (\ref{CurvedVotexA}) is functionally
dependent of the system (\ref{taucero},\ref{tautheta},\ref{taurho},\ref%
{conservation}).

We will consider the potential $V(f)$\ also as a variable of the system of
differential equations (\ref{taucero},\ref{tautheta},\ref{taurho}) jointly
with (\ref{CurvedVotexA}), That is a system of 4 non linear differential
equations with 4 variables $(m,l,f,V)$. With straightforward combinations,
the complete system could be written as:%
\begin{eqnarray}
f^{\prime \prime }+(4m+l)f^{\prime }-\frac{d}{df}V &=&0,  \label{system1} \\
l^{\prime }+(4m+l)l &=&-\frac{1}{2}V,  \label{system2} \\
m^{\prime }+(4m+l)m &=&-\frac{1}{2}V,  \label{system3} \\
m^{\prime }+m^{2}-ml &=&-\frac{1}{4}(f^{\prime })^{2}.  \label{system4}
\end{eqnarray}%
Equations (\ref{system2}) and (\ref{system3}) implies%
\begin{equation}
m=l,  \label{ml}
\end{equation}%
that can be obtained exactly considering\ a product of function solution: $%
l=q(r)\ m,$ due that the only consistent solution is \ $q(r)=1.$

So the system reduces to 3 equations:
\begin{eqnarray}
f^{\prime \prime }+5m.f^{\prime }-\frac{d}{df}V &=&0,  \label{Efe} \\
m^{\prime }+5m^{2} &=&-\frac{1}{2}V,  \label{m} \\
m^{\prime } &=&-\frac{1}{4}(f^{\prime })^{2}.  \label{mprima}
\end{eqnarray}%
as before for a given $V(f)$ the system (\ref{Efe},\ref{m},\ref{mprima}) is
not longer independent, and always one of the equations can be obtained from
the other two. But still solving the system is not an easy task, due to its
coupled non linear nature.

We will follow here the approach developed in \cite{grg} and \cite{prd},
instead of try to solve the system for a given $V(f),$ we will give a probe
function $f(r)$ that accomplishes the boundary conditions, and decouples the
system.

From (\ref{mprima}) it is easy to obtain $m$ as:%
\begin{equation}
m=-\frac{1}{4}\int dr(f^{\prime })^{2}-k_{RS},  \label{integralM}
\end{equation}%
where $k_{RS}$ is an additional Randall Sundrum constant warp factor. We
could obtain the potential $V(r)$ as a function of $r$ either from (\ref{m})
or (\ref{Efe}), consistency is guaranteed from conservation equation (\ref%
{conservation}) using (\ref{ml}).

In order to obtain the interaction potential $V(f)$ we must solve $r(f)$
inverting the probe function $f(r)$ so:
\begin{equation}
V(f)=V(r(f)),  \label{Vf}
\end{equation}%
Of course a solution to be physically acceptable must have a stable
potential with spontaneous breaking of symmetry, as we will show immediately.

Equations (\ref{Efe},\ref{m},\ref{mprima}) are very similar to that of 5D
domain walls in \cite{grg}\cite{Guerrero}, so we will try:
\begin{equation}
f=f_{0}\arctan \left( \sinh \ \beta r/\delta \right) ,\qquad f_{0}=2\sqrt{%
\delta },  \label{ffo}
\end{equation}%
using (\ref{integralM}) we get
\begin{equation}
m=-\beta \tanh \left( \beta r/\delta \right) ,\text{ \ where we set\ \ \ }%
k_{RS}=0,  \label{mchica}
\end{equation}

That solution corresponds to the metric "warped" factor
\begin{equation}
M(r)=\cosh ^{-\delta }(\beta r/\delta ),\qquad \delta >0,\qquad \beta >0.
\label{Mgrande}
\end{equation}

From (\ref{m}) we obtain the potential $V=V(r)$ as:
\begin{equation}
V(r)=\frac{2\beta ^{2}}{\delta }[(1+5\delta ){sech}^{2}(\beta r/\delta
)-5\delta ],  \label{VdeR}
\end{equation}%
and solving $r$ as a function of $f$ by means of (\ref{ffo}) we have
\begin{equation}
\cos^{2}(\frac{f}{f_{o}})= sech ^{2}(\frac{\beta r}{\delta }),
\end{equation}

Finally the potential is
\begin{equation}
V(f)=\frac{2\beta ^{2}}{\delta }[(1+5\delta )\cos ^{2}(f/f_{o})-5\delta ].
\label{VdeF}
\end{equation}

\begin{figure}[h]
\begin{minipage}[h]{0.3\linewidth}
\includegraphics[width=5cm,angle=0]{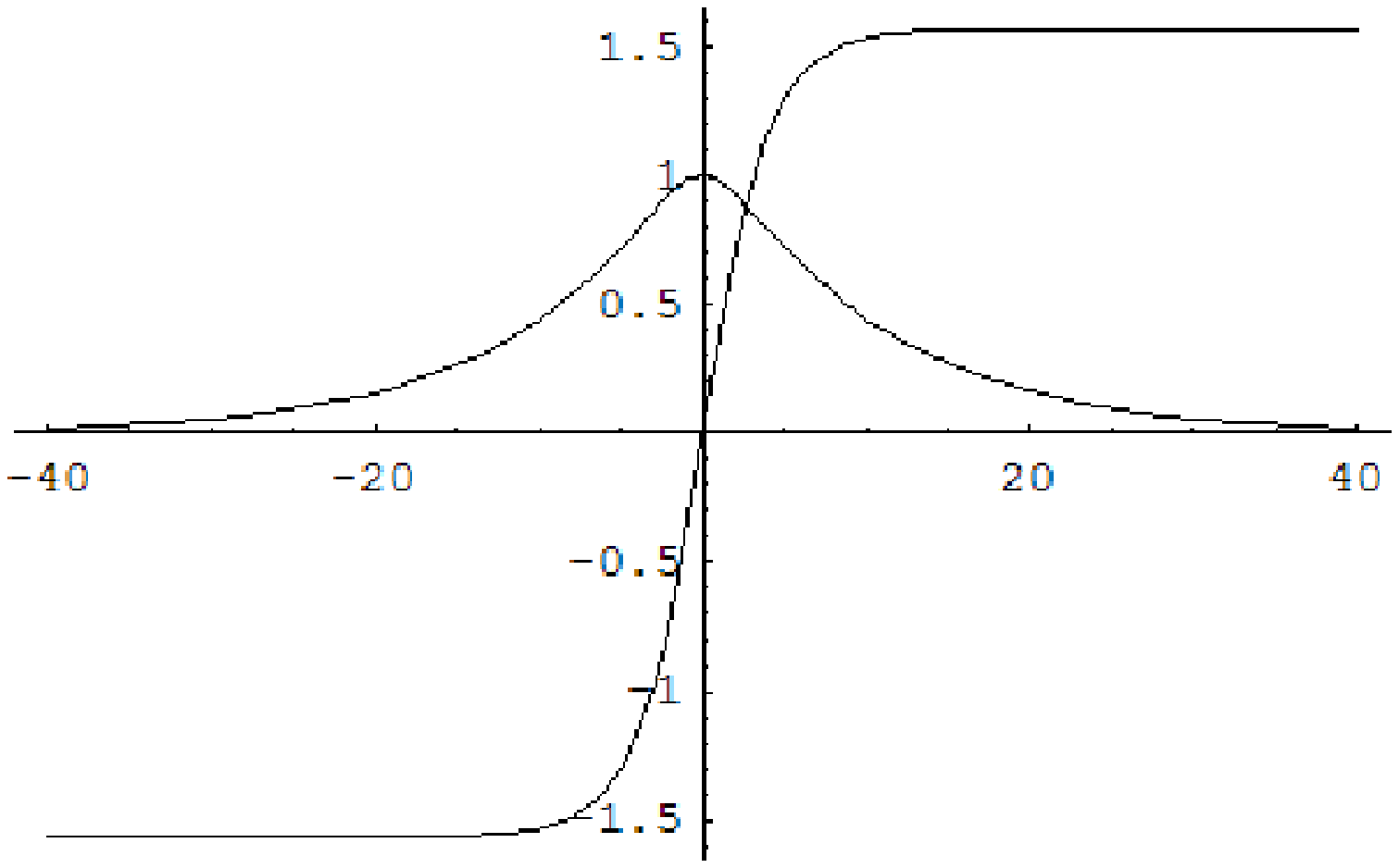}
\end{minipage}
\begin{minipage}[h]{0.3\linewidth}
\includegraphics[width=5cm,angle=0]{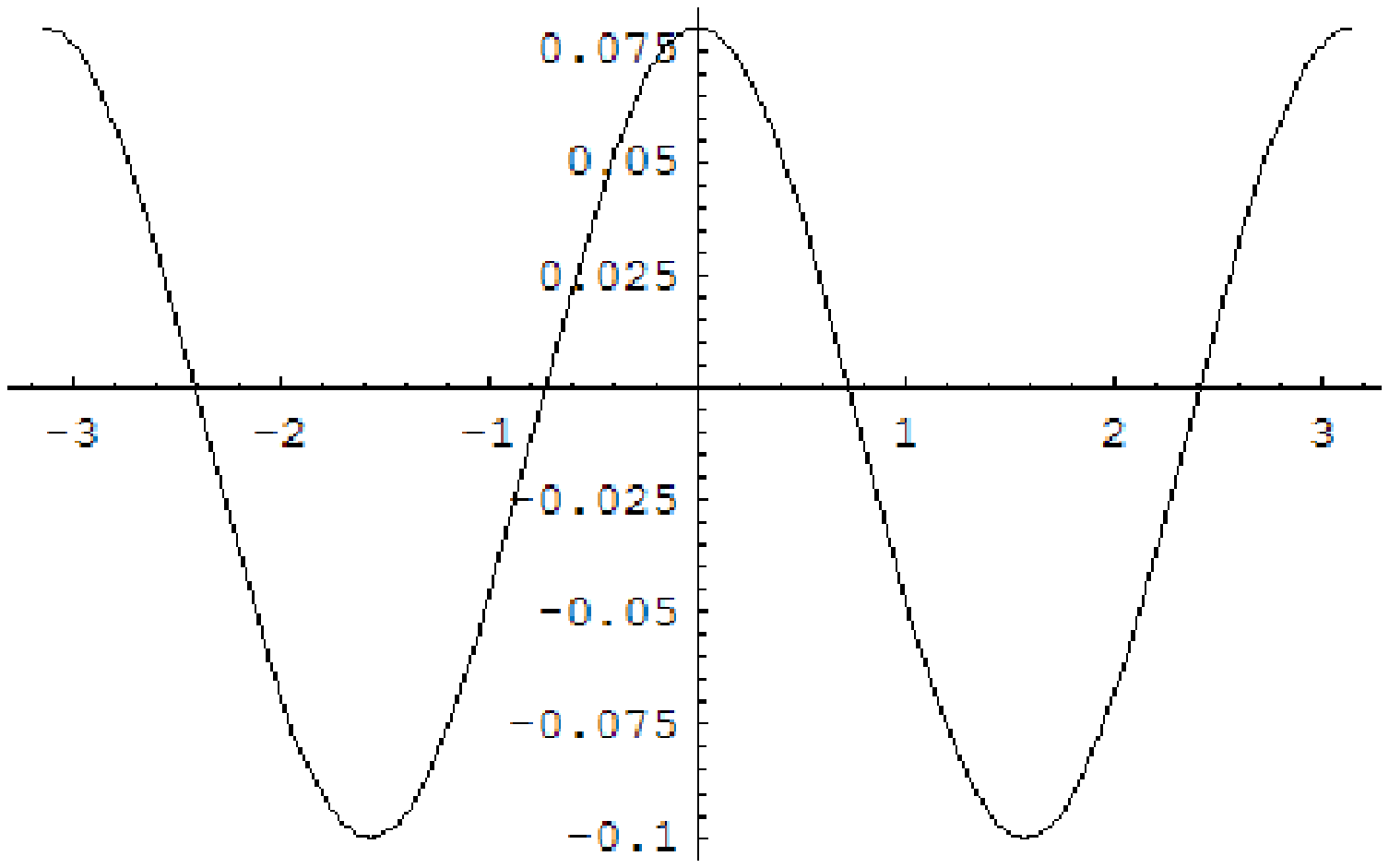}
\end{minipage}
\begin{minipage}[h]{0.3\linewidth}
\includegraphics[width=5cm,angle=0]{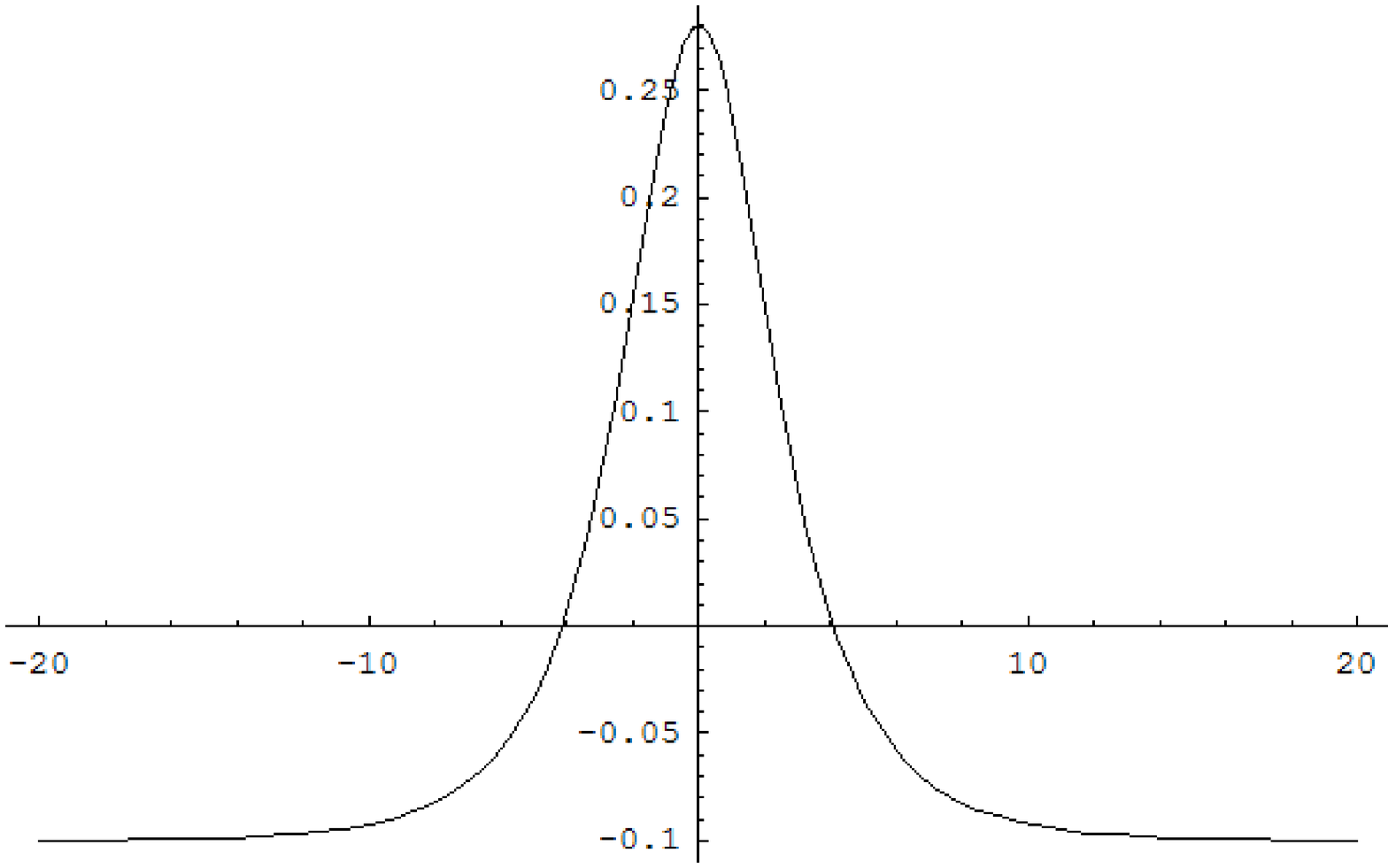}
\end{minipage}
\caption{Graphic of the metric factor M and scalar field (left), the
potential (center), and the energy density (right). fo=1, $\protect\delta %
=1/4,$ $\protect\beta =0.1$}
\label{MetricPotential}
\end{figure}

This is an stable potential, with spontaneous symmetry breaking minima,
where the scalar field interpolates smoothly between$\ \pm \pi \sqrt{\delta }%
,$ that in fact is the same point due to the odd parity of $r$coordinate.
This is an AdS spacetime with cosmological constant $\Lambda =10\beta ^{2}$
as could be seen from and Fig.\ref{MetricPotential}.

Here we will present another two new solutions, these are found by setting:

\begin{eqnarray}
f &=&f_{0}\tanh \left( a\ r\right) ,  \label{ftanh1} \\
\text{\ }f &=&f_{0}\arctan \left( a\ r\right) ,  \label{ftanh2}
\end{eqnarray}%
using (\ref{integralM}) we get $m(r)$, and setting $k_{RS}=0$ as before, we
obtain the metric "warped" factors, respectively
\begin{eqnarray}
M(r) &=&\exp [\frac{1}{24}f_{0}^{2}\ sech(a\ r)]\cosh ^{-f_{o}^{2}/6}(a\ r),
\label{Mtan1} \\
M(r) &=&\exp [-\frac{a}{2}f_{0}^{2}\ r\arctan (a\ r)],\text{\ \ }
\label{Mtan2}
\end{eqnarray}

From (\ref{m}) we obtain the potential $V=V(r)$ and solving r as a function
of f as before we get the potentials, respectively
\begin{eqnarray}
V(f) &=&\frac{a^{2}}{2}[(f^{2}-f_{o}^{2})^{2}-\frac{5}{9}%
f^{2}(f^{2}-3f_{o}^{2})],  \label{VdF2} \\
V(f) &=&\frac{a^{2}f_{o}^{2}}{2}[\cos ^{4}(f/f_{o})-\frac{5}{16}%
f_{o}^{2}[f/f_{0}+\frac{1}{2}\sin (f/f_{o})]^{2}  \label{Vdf3}
\end{eqnarray}

The potentials (\ref{VdF2}) and (\ref{Vdf3}) with the metrics warped factors
(\ref{Mtan1}) and (\ref{Mtan2}), that corresponds to the scalar "kink"
fields (\ref{ftanh1}) and (\ref{ftanh2}) are very similar in shape to the
former solution (\ref{VdeF}),(\ref{Mgrande}) and (\ref{ffo}) plotted in Fig.%
\ref{MetricPotential}. As before the potentials presents symmetry breaking
minima, and the scalar field is also an interpolating soliton between those
minima in AdS space time with cosmological constant $\Lambda =-\frac{5}{18}%
a^{2}f_{0}^{4}$\ \ and $\ \Lambda =-\frac{5}{128}a^{2}\pi ^{2}f_{0}^{4}$,
respectively.

Remarkably these string-vortex solutions are exact solutions with integer
winding number and finite 4 dimensional energy density, integrating $\tau
_{o}$ (\ref{Trho}) over the extra dimensions due to the decaying warping
factor $M$ (\ref{Mgrande},\ref{Mtan1},\ref{Mtan2}), as could be saw in Fig.%
\ref{MetricPotential}. Note that the potential (\ref{VdF2}) is a quartic
polynomial potential\ and could be considered as the first known exact
solution to the Abelian Higgs model with a mexican hat potential.

Until now, almost all known solutions in 6D curved space-time were
asymptotical or numerical \cite{RSVortex}\cite{Shaposhnikov} with the only
exception of the one in \cite{belga} and the critical string limit \cite%
{Bogomoln'yi}\cite{Vega}\cite{Preskill}. If we set $f=1$ in (\ref{Efe},\ref%
{m},\ref{mprima}) we obtain $M=\exp [-\beta r]$ in concordance with the
result in \cite{belga}. In general the kind of solutions we had found, are
very important to establish the possibility of confinement, quantization and
stability of other fields as fermions, photons, gravitons, etc., opening the
possibility to further developments.

\begin{figure}[h]
\begin{minipage}[h]{0.50\linewidth}
\includegraphics[width=5cm]{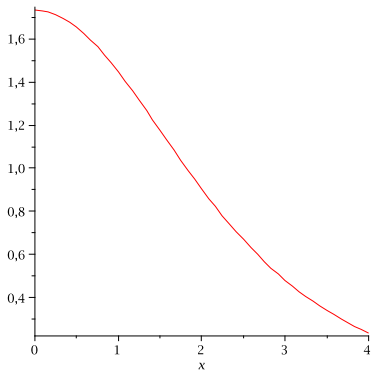}
\end{minipage}
\caption{ Massless wavefunction for $\protect\upsilon =0,$ $\protect\beta =1$%
}
\label{masslessplot}
\end{figure}


\section{Localized gravity on the topological string-vortex}


Next, we will show that these new vortices can confine gravity. In the
previous section we found an exact solution to the Einstein equations for
the metric (\ref{metric},\ref{ml},\ref{Mgrande}) for the case that the gauge
field and scalar field are given by (\ref{F},\ref{Atheta},\ref{Ar},\ref{Amu}%
\ref{P},\ref{ffo}) with the non trivial potential (\ref{VdeF}). We will
study linearized spin 2 metric fluctuations from the metric (\ref{Mgrande})
neglecting graviscalar and graviphotons modes.
\begin{equation}
h_{\tau \nu }(x,r,\theta )=\eta _{\tau \nu }e^{ip\cdot x}\sum_{\kappa ,\mu
}\phi _{\upsilon }(r)e^{i\kappa \theta },  \label{fluctuation}
\end{equation}%
where $x$ stands for $\ x^{\mu }$ the cordinates on the 4 dimensional branes
and $(r,\theta )$ for \ the polar coordinates of the bulk. These massive
modes (\ref{fluctuation}) obeys the Laplace Beltrami operator, and
straightforward the radial modes satisfy \cite{Sov.Phys}%
\begin{equation}
-\frac{1}{M^{2}L}\frac{\partial }{\partial r}\left[ \sqrt{-g}\frac{\partial
}{\partial r}(\phi _{\upsilon })\right] =\upsilon ^{2}\phi _{\upsilon },
\label{wave}
\end{equation}%
where $\upsilon ^{2}=m_{o}^{2}-\left( \frac{M}{L}\right) ^{2}(\kappa )^{2}$
represents the mass of mode $\phi _{\upsilon }$ with square 4 dimensional
momentum $p_{\mu }p^{\mu }=m_{o}^{2}$. The mass term takes into account the
contributions from orbital angular momentum $\kappa $, as in \cite%
{Shaposhnikov} while $\upsilon ^{2}>0$ due it takes into account the
contribution of the (plus) radial momentum square.

Upon substitution of (\ref{ml},\ref{Mgrande}) we get the equation for
massive metric fluctuations:%
\begin{equation}
-\phi _{\upsilon }^{\prime \prime }+5\beta \tanh (\beta r/\delta )\phi
_{\upsilon }^{\prime }-\upsilon ^{2}\cosh (\beta r/\delta )\phi _{\upsilon
}=0,  \label{massive}
\end{equation}%
when $\upsilon ^{2}=0$ the equation (\ref{massive}) reduces to:
\begin{equation}
\phi _{o}^{\prime }\cosh ^{-5\delta }(\beta r/\delta )=k_{1},
\label{massless}
\end{equation}%
integrating (\ref{massless}), we obtain the massless mode
\begin{equation}
\phi _{o}=k_{o}+k_{1}\int dr\cosh ^{5\delta }(\beta r/\delta ).  \label{Phi0}
\end{equation}%
with integrations constants $\ k_{o}$ \ and\ $\ k_{1}$.

As $\cosh ^{5\delta }[\beta r/\delta ]$ is monotonous growing, we must fix $%
k_{1}=0$, in order to render $\phi _{o}$ \ normalizable. That is consistent
with the boundary conditions%
\begin{equation}
\phi _{\upsilon }^{\prime }(0)=\phi _{\upsilon }^{\prime }(\infty )=0,
\end{equation}%
that allows (\ref{wave}) to be a self adjoint Sturm-Liouville well possed
problem \cite{Shaposhnikov}.

The ortonormalization condition to be satisfied by $\phi _{\upsilon }$ is%
\begin{eqnarray}
\int_{0}^{\infty }dr\ M^{2}L\ \phi _{m}^{\ast }(r)\phi _{\upsilon }(r)
&=&\delta _{m\upsilon }, \\
\int_{0}^{\infty }dr\cosh ^{-3\delta }(\beta r/\delta )\ \phi _{m}^{\ast
}.\phi _{\upsilon } &=&\delta _{m\upsilon },  \notag
\end{eqnarray}%
so the equivalent wavefunction in 1-dimensional quantum mechanics is

\begin{equation}
\psi _{\upsilon }(r)=\cosh ^{-3\delta /2}(\beta r/\delta )\ \phi _{\upsilon
}(r).  \label{massivewave}
\end{equation}

Finally, the massless normalized equivalent wavefunction is given by:%
\begin{equation}
\psi _{o}(r)=k_{o}\cosh ^{-3\delta /2}(\beta r/\delta ).
\label{masslesswave}
\end{equation}%
This function is strongly decaying (see Fig.\ref{masslessplot}), so we
conclude that the massless, spin two, gravitation mode is localized on the
3-brane and strongly concentrated around $r=0$ as we expected for a RS
scenario \cite{RSVortex}\cite{Shaposhnikov}.

There is another approach to study the localization of gravity modes around
the vortex \cite{Castillo-Felisola}\cite{Wang2002} in an exact way, the
equation (\ref{massive}) could be written as%
\begin{equation}
-\phi _{\upsilon }^{\prime \prime }-5\frac{M^{\prime }}{M}\phi _{\upsilon
}^{\prime }=\frac{\mu ^{2}}{M^{2}}\phi _{\upsilon },
\end{equation}%
that could be converted with the change of variables%
\begin{equation}
u_{\upsilon }(r)=M^{5/2}\phi _{\upsilon }(r),  \label{umu}
\end{equation}%
into a 1 dimensional stationary Schr\"{o}dinger equation with zero eing\"{e}%
nvalue:
\begin{equation}
(-\partial _{r}^{2}+V_{QM})\ u_{\upsilon }=0,  \label{eq.Schrodinger}
\end{equation}%
here
\begin{equation}
V_{QM}=\frac{15}{4}\left( \frac{M^{\prime }}{M}\right) ^{2}+\frac{5}{2}%
\left( \frac{M^{\prime \prime }}{M}\right) -\left( \frac{\upsilon }{M}%
\right) ^{2}.  \label{Vqm Eq}
\end{equation}%
Note by squaring and integrating (\ref{umu}) that if $u_{\upsilon }$ is a
bounded state, then $\phi _{\upsilon }$ is also bounded.

\begin{figure}[b]
\begin{minipage}[b]{0.30\linewidth}
\includegraphics[width=4cm,angle=0]{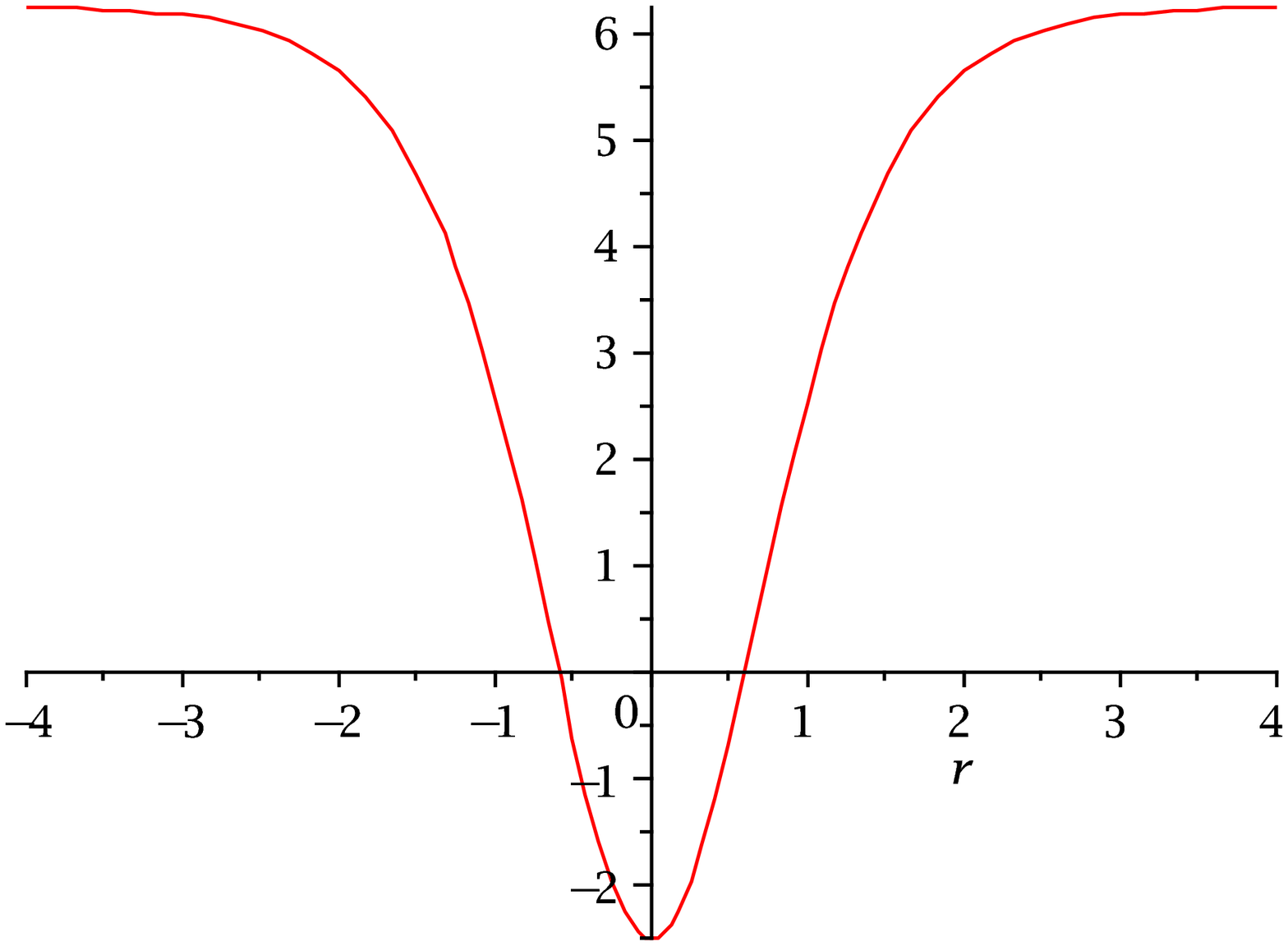}
\end{minipage}%
\begin{minipage}[b]{0.30\linewidth}
\includegraphics[width=4cm]{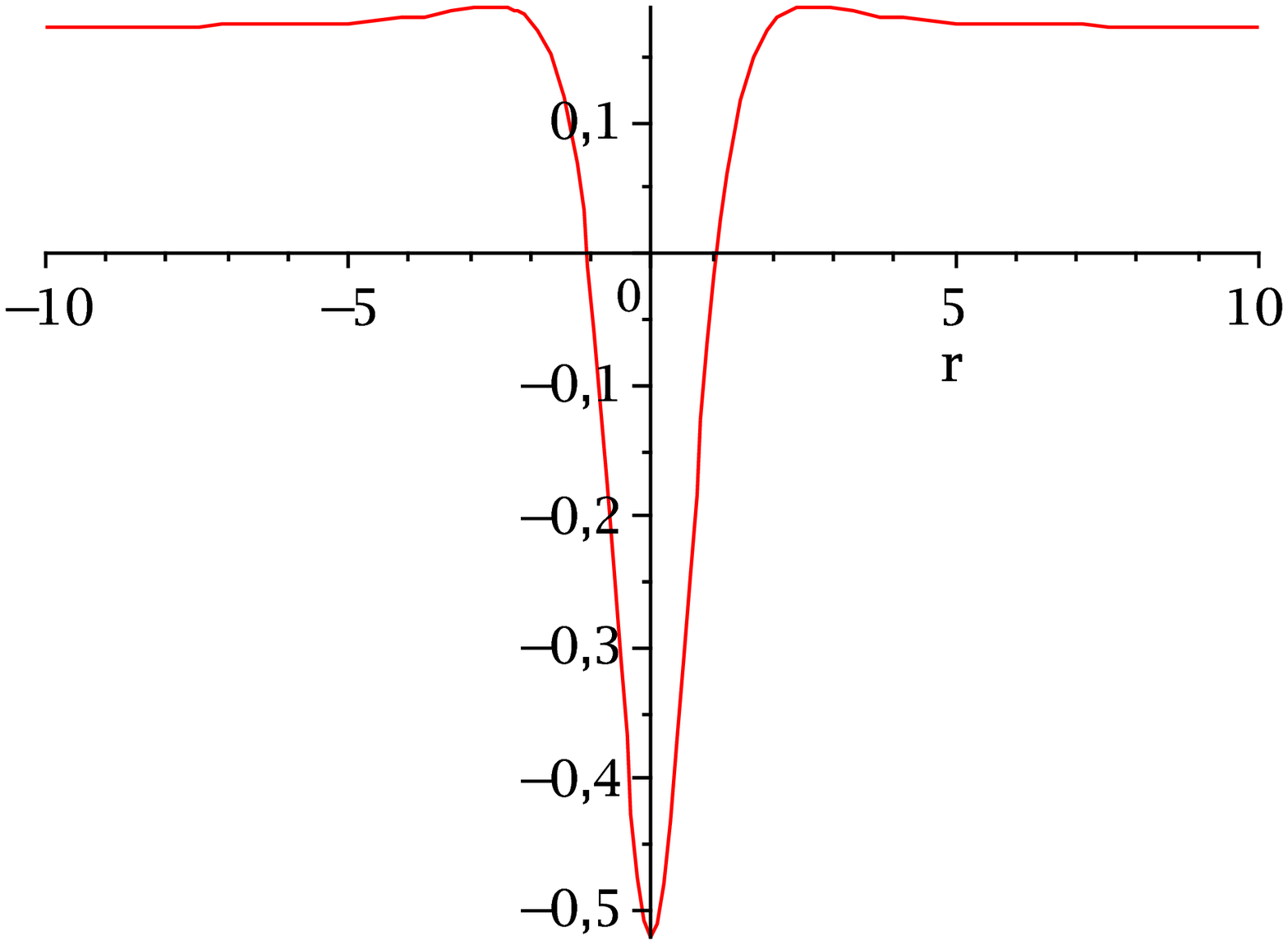}
\end{minipage}%
\begin{minipage}[b]{0.30\linewidth}
\includegraphics[width=4cm]{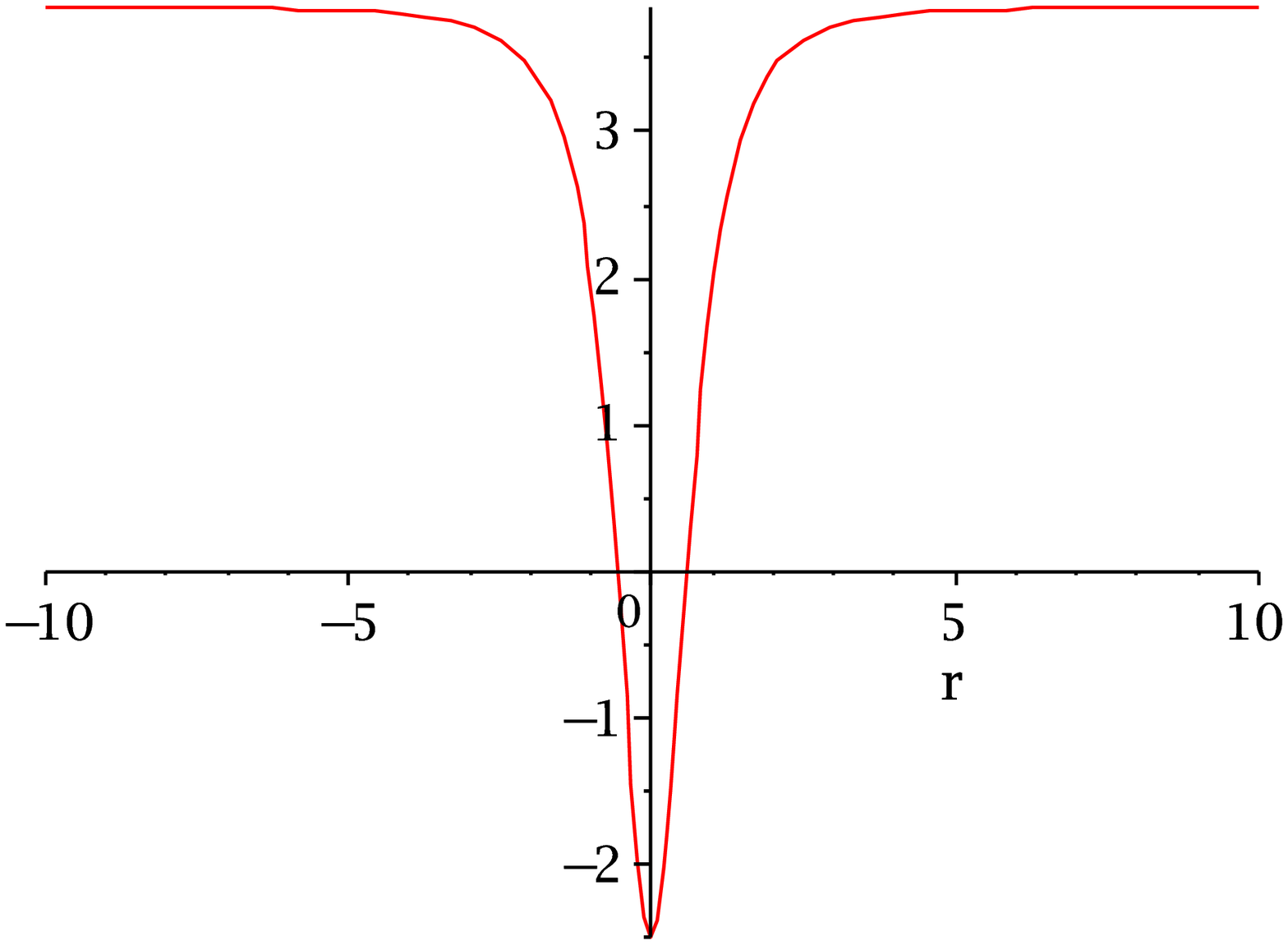}
\end{minipage}
\caption{Graphics of the potential $V_{QM}$ for the massless mode of the
vortices with M given by (\protect\ref{Mgrande}) at the left ($\protect\beta %
=1$, $\protect\delta =1$). At the center by (\protect\ref{Mtan1}) and at the
right by (\protect\ref{Mtan2}) (fo=1, a=1).}
\label{VqmPlot}
\end{figure}

For the case of solution(\ref{Mgrande}), the quantum potential for massive
states is:%
\begin{equation}
V_{QM}=-\upsilon ^{2}\cosh ^{2\delta }(\beta r/\delta )+\frac{5\beta ^{2}}{2}%
\left[ \frac{5}{2}\tanh ^{2}(\beta r/\delta )-\frac{1}{\delta }%
sech^{2}(\beta r/\delta )\right] .  \label{VQM}
\end{equation}

\begin{figure}[t]
\begin{minipage}[t]{0.30\linewidth}
\includegraphics[width=4cm,angle=0]{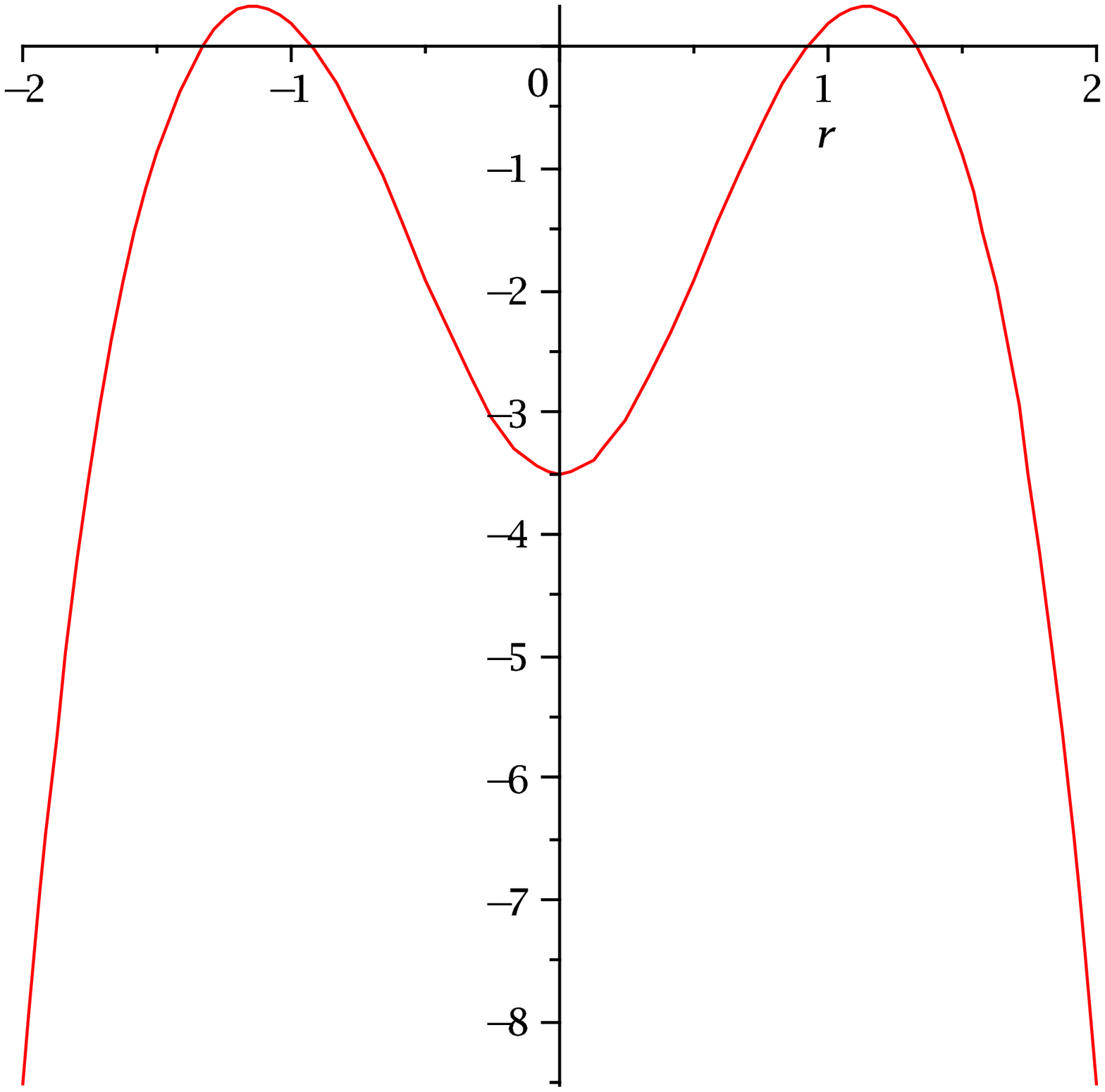}
\end{minipage}%
\begin{minipage}[t]{0.30\linewidth}
\includegraphics[width=5cm]{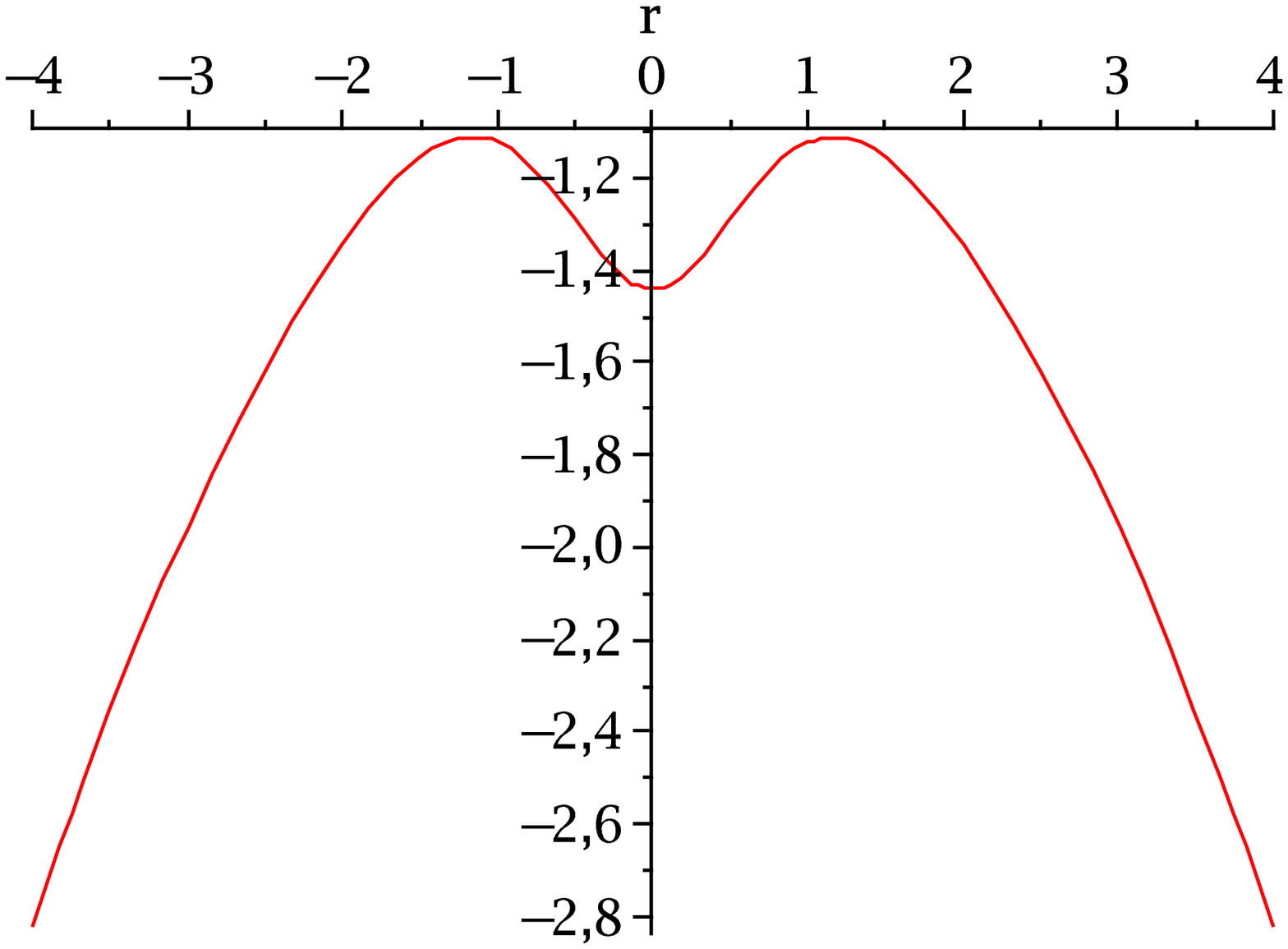}
\end{minipage}%
\begin{minipage}[t]{0.30\linewidth}
\includegraphics[width=5cm]{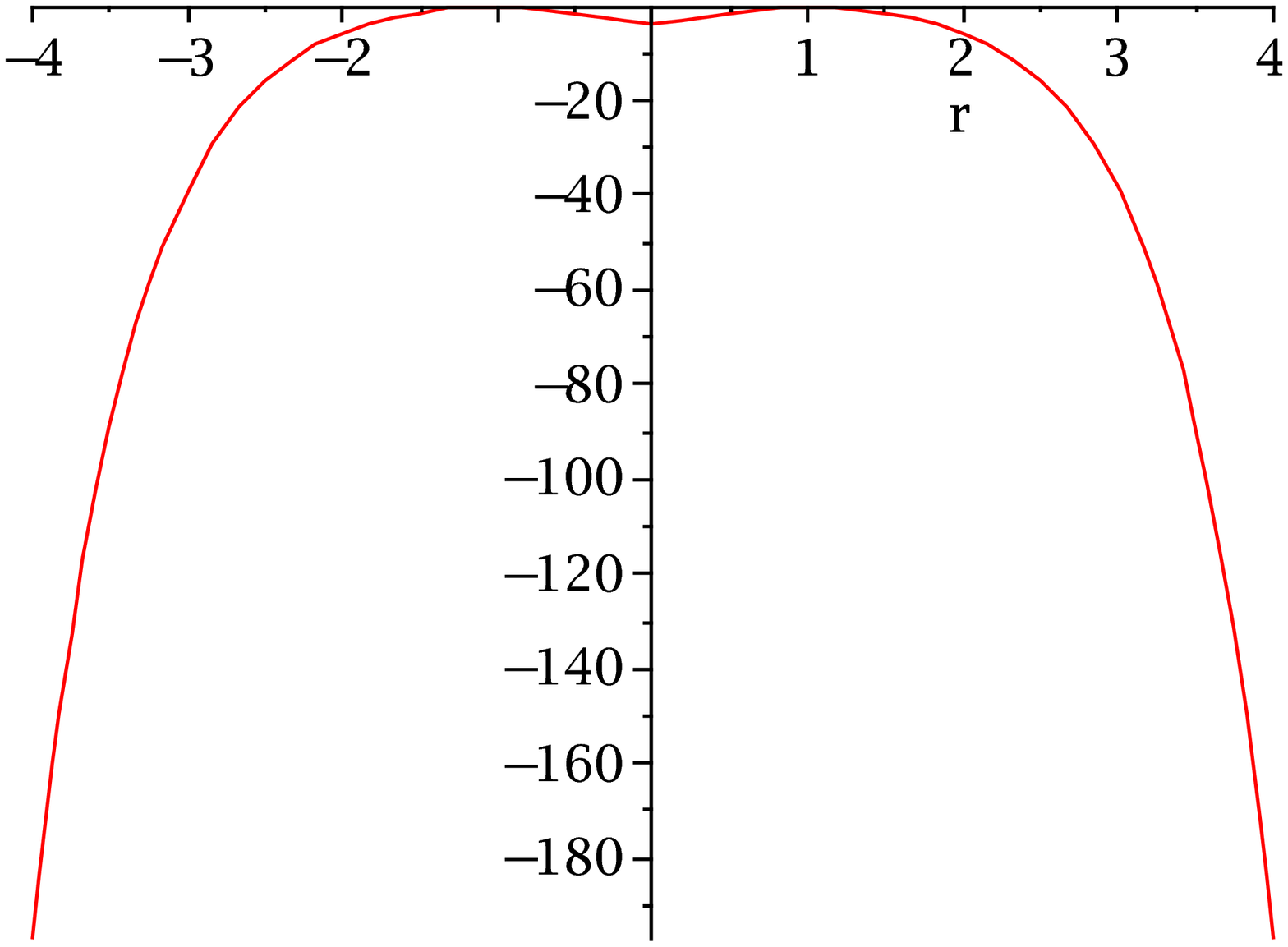}
\end{minipage}
\caption{Graphics of the potential $V_{QM}$ for the massive modes $\protect%
\upsilon =1$\ of the vortices with M given by (\protect\ref{Mgrande}) at
left for $\protect\beta =1$ \ and $\protect\delta =1$. At the center by (%
\protect\ref{Mtan1}) and at the right (\protect\ref{Mtan2}) for fo=1 and
a=1. }
\label{VqmMass}
\end{figure}

Due to the negative term in the first term in the quantum potential (\ref%
{VQM}), there will be not bounded states except for $\upsilon =0$, as could
be seen from Fig.\ref{VqmPlot}. When $\upsilon \neq 0,$ the potential is not
bounded from below, as could be observed from Fig.\ref{VqmMass} so the
potential well is unstable and the massive states will easily tunnel outside
the well. Therefore only the massless state will be bounded.

In a similar way the equivalent quantum mechanical potential for the other
two vortex solutions (\ref{ftanh1},\ref{Mtan1}) and (\ref{ftanh2},\ref{Mtan2}%
) could easily be obtained from equation (\ref{Vqm Eq}), we can again
observe in Fig.\ref{VqmPlot} that when $\upsilon =0$, \ there exist a
potential well that allows the existence of a confined massless gravity
ground state, but when $\upsilon \neq 0$, \ the potential has a negative
exponential growing, as could be seen in Fig.\ref{VqmMass}, so is not
bounded from below and massive bounded states are forbidden.

\section{Massive gravity modes and newtonian limit of gravity.}

In order to compute the newtonian limit of gravity we need exact solutions
to massive modes equation (\ref{massive}). This equation can be solved for
particular values of $\delta $ \ in term of Heun confluent $H_{C}$ functions
\cite{Heun}:

For $\ \delta =1$ we get for $\phi _{\upsilon }(r)$:%
\begin{equation}
C_{1\ }H_{C}[0,-\frac{{\ 1}}{{\ 2}},-3,-\frac{{\ }\upsilon ^{{\ 2}}}{{\
4\beta }^{{\ 2}}},\frac{{\ 4\beta }^{{\ 2}}{+}\upsilon ^{{\ 2}}}{{\ 4\beta }%
^{{\ 2}}},-\sinh ^{2}(\beta r)]+C_{2}\ \sinh (\beta r)\ H_{C}[0,\frac{{\ 1}}{%
{\ 2}},-3,-\frac{{\ }\upsilon ^{{\ 2}}}{{\ 4\beta }^{{\ 2}}},\frac{{\ 4\beta
}^{{\ 2}}{+}\upsilon ^{{\ 2}}}{{\ 4\beta }^{{\ 2}}},-\sinh ^{2}(\beta r)],
\label{Heun}
\end{equation}

while for $\ \delta =2$ we get for $\phi _{\mu }(r)$:%
\begin{eqnarray}
&&C_{1}\ e^{\frac{-i\upsilon }{\beta }\sinh ^{2}(\frac{{\ \beta }r}{2})}\
H_{C}[\frac{{\ 2i}\upsilon }{{\ \beta }}{\ ,-}\frac{{\ 1}}{{\ 2}}{\ ,-}\frac{%
{\ 11}}{{\ 2}}{\ ,-}\frac{{\ }\upsilon ^{{\ 2}}}{{\ \beta }^{{\ 2}}}{\ ,}%
\frac{{\ 13\beta }^{{\ 2}}{+8}\upsilon ^{{\ 2}}}{{\ 8\beta }^{{\ 2}}}{\ ,-}%
\sinh ^{2}{\ (}\frac{\beta r}{2}{\ )}]+ \\
&&\qquad \qquad C_{2}\sinh (\frac{\beta r}{2})\ e^{\frac{-i\upsilon }{{\
\beta }}\sinh ^{2}(\frac{{\ \beta }r}{2})}\ H_{C}[\frac{{\ 2i}\upsilon }{{\
\beta }}{\ ,}\frac{{\ 1}}{{\ 2}}{\ ,-}\frac{{\ 11}}{{\ 2}}{\ ,-}\frac{{\ }%
\upsilon ^{{\ 2}}}{{\ \beta }^{{\ 2}}}{\ ,\frac{13\beta ^{2}+8\upsilon ^{2}}{%
8\beta ^{2}},}-\sinh ^{2}(\frac{\beta r}{2})].  \notag
\end{eqnarray}

A plot of the massive wavefunction (\ref{massivewave}) is given in Fig. (\ref%
{waveplot}) when $\delta =1,$ $\upsilon =1$ and $\beta =1$ using (\ref{Heun}%
). In that figure its oscillating growing nature could be observed.

Although general wavefunction solutions are rather cumbersome, what really
matters is the asymptotic behavior of the massive modes. Far from the vortex
core or equivalently for $\delta \rightarrow 0,$ that correspond to the thin
string limit, we can approximate (\ref{Mgrande}) by%
\begin{equation}
M=\cosh ^{-\delta }(\beta r/\delta )\cong (\frac{{\ 1}}{{\ 2}})^{\delta
}.e^{-\beta r},  \label{aproximation}
\end{equation}%
for the other two vortices (\ref{ftanh1},\ref{Mtan1}) and (\ref{ftanh2},\ref%
{Mtan2}), we were not able to found exact solutions to equation (\ref{wave}%
), but we can obtain the same asymptotic limit (\ref{aproximation}) for the
second exact solution (\ref{Mtan1}) with $\delta =f_{o}^{2}/6$ \ and $\beta
=a\ f_{o}^{2}/6$ and for the third exact solution (\ref{Mtan2}) if we set $%
\delta =1/a$ \ and $\beta =a\pi \ f_{o}^{2}/4$ . So the massive wavefunction
(\ref{massive}) could be approximate for all these three vortex solutions by%
\begin{equation}
\psi _{\upsilon }(r)\cong e^{-\frac{3}{2}\beta r}\ \phi _{\upsilon }(r),
\label{fiMiu}
\end{equation}%
and the localized zero mode is given by%
\begin{equation}
\psi _{o}(r)\cong \sqrt{3\beta }e^{-\frac{3}{2}\beta r}\ ,
\label{groundstate}
\end{equation}

\begin{figure}[h]
\begin{minipage}[h]{0.50\linewidth}
\includegraphics[width=5cm]{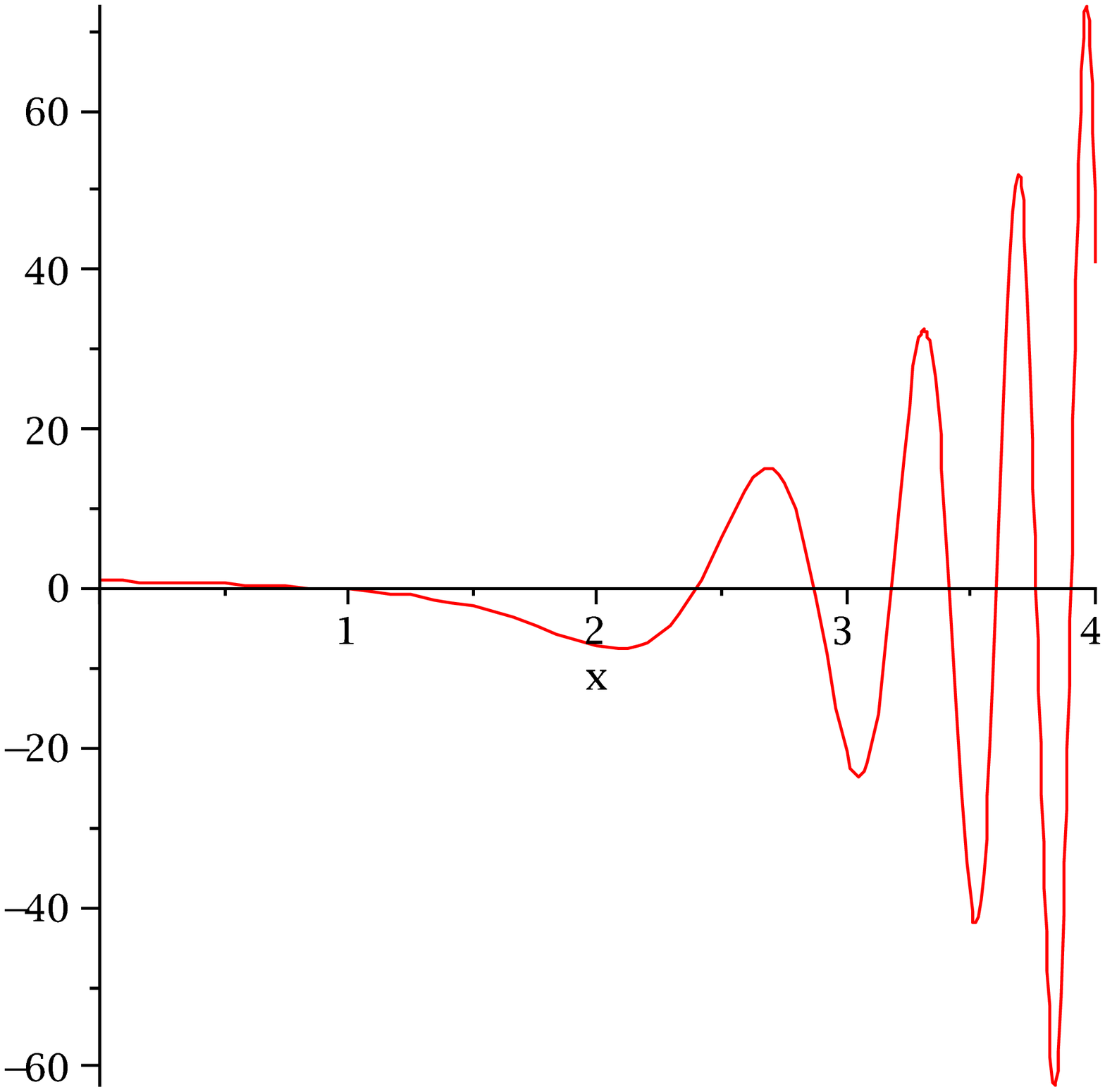}
\end{minipage}
\caption{Massive wavefunction with$\ \ \protect\upsilon =1,$ $\protect\beta %
=1$, for Eq.(\protect\ref{Heun}, \protect\ref{massivewave})}
\label{waveplot}
\end{figure}

The massless zero mode is then localized in the vicinity of $r=0$, that
means that massless gravitons, that mediates "normal" 4D gravity, could only
be found near the 3 brane where the known universe is located, decaying
exponentially in number when when $r$ is increased, while far from the
3-brane only could be found as seen from Fig.\ref{waveplot} showing the
behavior of the massive wave functions.

Using the approximation (\ref{aproximation}) and (\ref{fiMiu}) into the
differential equation (\ref{massive}), and taking the limit $\beta r/\delta
\rightarrow \infty $ for which $\tanh (\beta r/\delta )\rightarrow 1,$ we
obtain%
\begin{equation*}
-\phi _{\upsilon }^{\prime \prime }+5\beta .\phi _{\upsilon }^{\prime
}-\upsilon ^{2}e^{2\beta r}\phi _{\upsilon }=0,
\end{equation*}%
which solution is given in term of bessel functions $J_{5/2}(\frac{\upsilon
}{\beta }e^{\beta r})$ and $Y_{5/2}(\frac{\upsilon }{\beta }e^{\beta r})$.
In exact concordance with the result in \cite{Shaposhnikov}.

The newtonian limit was obtained in \cite{Shaposhnikov}, using a radial
cutoff approach calculation as:%
\begin{equation}
V\cong \frac{{\ 1}}{{\ m}_{{\ pl}}^{{\ 2}}}\frac{m_{1}.m_{2}}{R}\left[ 1+%
\frac{{\ 4}}{{\ 3\pi \beta }^{{3}}}\frac{1}{R^{3}}+...\right] ,
\label{newtonian}
\end{equation}%
where $m_{pl}^{2}=\frac{2\pi }{3\beta }(M_{6})^{4}$ is the 4 dimensional
Planck mass, and $R$ is the distance between two particles of masses $m_{1}$
and $m_{2}$ on the 3-brane. Therefore the $\frac{1}{R^{3}}$\ correction to
the newtonian gravity decays stronger than $\frac{1}{R^{2}}$\ correction in
5 dimensional RS domain walls \cite{RS2} due to the additional dimension and
in the thin string limit $\beta \rightarrow \infty ,$ the correction to the
newtonian potential vanishes.

\section{Summary and outlook}


We consider the central result of this paper, is to have found exact
topological solutions embedded in curved 6D spacetime that localize gravity.
These solutions, were obtained following the procedure developed in \cite%
{grg} and \cite{prd} to obtain solutions to Einstein equations with proposed
scalars kinks sources for suitable scalar potentials. These topological
solutions corresponds to different electromagnetic field vacua and are, in
fact, string-vortices in 6D with non trivial winding number.

We found three new exact 6D vortex solutions, for three different kinks,
that renders adequate warped metric factors and potentials with degenerate
minima that allows spontaneous symmetry breaking of the theory. The first
solution (\ref{ffo}), (\ref{Mgrande}) and (\ref{VdeF}) is the vortex version
of the domain wall solution in \cite{Guerrero} where $\delta $ could be seen
as the string thickness and $\beta $ as a parameter depending on the
cosmological constant. The second solution (\ref{ftanh1}), (\ref{Mtan1}) and
(\ref{VdF2}) has a quartic polynomial potential that is almost identical to
the mexican hat potential. The third solution (\ref{ftanh2}), (\ref{Mtan2})
and (\ref{Vdf3}) combine polynomial and not polynomial terms in the
potential. All the new solutions presented here, are very similar to the
domain walls, but the new solutions includes axial symmetry in 6 dimension
and a gauge background with integer winding number, that is absent in 5
dimensional domain walls.

As expected for a RS brane world, we have shown that the zero mode (\ref%
{Phi0})\ of the linearized gravity spectrum is localized on the 3-brane
while the massive modes (\ref{massivewave},\ref{Heun}) are not bounded at
all. The results of section III were obtained for the 6D topological vortex
solutions without approximations. The newtonian limit was also obtained in
complete concordance with known results \cite{Shaposhnikov}.

Using exact solutions many interesting properties of 6D brane world could be
study. If we use these topological vortex solutions as background, we can
proceed to study the quantization and stability under fluctuations, it is
also possible to study the confinement of electromagnetic spin 1 and other
gauge or fermionic fields in an exact way. These subjects are currently
under research \cite{preparation}.


\section*{Acknowledgments}

This work was supported by CDCHT-UCLA under project 005-CT-2007.


\bigskip

\end{document}